\documentclass[journal=ancac3,manuscript=article]{achemso}

\usepackage{chemformula} % Formula subscripts using \ch{}
\usepackage[T1]{fontenc} % Use modern font encodings
\usepackage{gensymb}

\author{Claudio Melis}
 \affiliation{Dipartimento di Fisica, Universit\`a  degli Studi di Cagliari, Cittadella Universitaria, I-09042 Monserrato (CA), Italy}
 \email{claudio.melis@dsf.unica.it}
\author{Giorgio Pia}
\affiliation{Dipartimento di Ingegneria Meccanica, Chimica, e dei Materiali, Universit\`a  degli Studi di Cagliari,via Marengo 2, I-09123 Cagliari, Italy}
\author{Elisa Sogne}
\affiliation{NABLA Lab, Biological and Environmental Sciences and Engineering (BESE) Division,King Abdullah University of Science and Technology (KAUST), Thuwal, Saudi Arabia}
\author{Andrea Falqui}
\affiliation{NABLA Lab, Biological and Environmental Sciences and Engineering (BESE) Division,King Abdullah University of Science and Technology (KAUST), Thuwal, Saudi Arabia}
\author{Stefano Giordano}
\affiliation{Univ. Lille, CNRS, Centrale Lille, Univ. Polytechnique Hauts-de-France, UMR 8520,- IEMN - Institut d'Electronique de Microélectronique et de Nanotechnologie, F-59000 Lille, France}
\author{Francesco Delogu}
\affiliation{Dipartimento di Ingegneria Meccanica, Chimica, e dei Materiali, Universit\`a  degli Studi di Cagliari,via Marengo 2, I-09123 Cagliari, Italy}
\author{Luciano Colombo}
\affiliation{Dipartimento di Fisica, Universit\`a  degli Studi di Cagliari, Cittadella Universitaria, I-09042 Monserrato (CA), Italy}
\title[An \textsf{achemso} demo]
  {Stiffening of nanoporous Au as a result of dislocation density increase upon characteristic length reduction}

\begin{document}


\begin{abstract}
Structure is the most distinctive feature of nanoporous metals. The intricate maze of rounded shapes, where ligaments and pores run after each other disorderly, strikes imagination no less than it imparts properties that, tuned by size effects, have no counterpart in the bulk form. Indisputably, nanoporous Au has been the absolute protagonist of the field of study, unveiling the disrupting potential of nanoporous metals in areas ranging from catalysis to energy and sensing. Here, we still focus on nanoporous Au, addressing the long-standing issue of mechanical properties in nanoporous metals. In particular, we investigate how Young's modulus changes with ligament size, being the porosity the same. Based on atomistic replicas generated starting from experimental tomographic evidence, atomistic simulations reveal that nanoporous Au stiffens as ligaments become finer, reproducing experimental findings obtained by nanoindentation of dealloyed samples. Ruled out surface stress effects, theoretical considerations relate stiffening to the dislocation density increase.
\end{abstract}

\section{Introduction}
Nanoporous metals (NPM) form a broad class of monolithic solids that are currently under intense scrutiny. With their architecture consisting of an elegant, and yet confusing, tangle of voids and solid \cite{lilleodden2018topological}, in the past 20 years these materials have never ceased to fascinate commoners and amaze scientists from very different backgrounds. Their history contributes to make the class of materials unique. Manufactured and utilized by ancient civilizations \cite{ogden1983goldsmithing,wittstock2012nanoporous}, time by time NP metals, and NP Au in particular, re-surface from the darkness of alchemical laboratories and the implications of corrosion phenomena \cite{reti1965parting,karpenko1988coins,berthelot1889introduction,evans1963introduction,forty1979corrosion}.

This is not surprising considering that the main synthetic route to NP metals is based on the selective dissolution of less noble metals from a parent alloy. Dealloying is, indeed, a relatively simple process that can be mastered by craftsmen unaware of chemistry fundamentals \cite{hunt1976oldest}. The alloy is exposed to an acidic aqueous solution and less noble elements are progressively removed by a free corrosion process \cite{erlebacher2001evolution,mccue2016dealloying,erlebacher2004atomistic}. At present, electrochemical methods are often applied to achieve a better control of the dissolution process \cite{mccue2016dealloying,zhang2012nanoporous}, but this leaves the key features of the corrosion process unaffected.

In short, dealloying relies upon the complex interplay between the chemistry of selective dissolution and the physics of surface rearrangement \cite{erlebacher2001evolution,mccue2016dealloying,erlebacher2004atomistic}. As quickly as atoms of less noble species are removed from the surface, the remaining atoms of most noble species diffuse from low-coordination surface sites to more stable ones, possibly exposing initially buried less noble atoms. A gradual change of characteristic lengths accompanies porosity formation on the nanometer scale.

The result of dealloying is an homogeneous, interpenetrating, bi-continuous disordered network of ligaments and pores \cite{lilleodden2018topological,mccue2016dealloying,fujita2008three}. The nature of metallic bond pervades harmoniously the whole architecture, ensuring that any reduction of characteristic lengths deeply impacts the physical and chemical properties of the NPM.

In this regard, NP Au represents the most typical example \cite{wittstock2012nanoporous}. Easy to obtain because of Au chemical nobility, it has provided the ideal model system to explore the structure-property relationships for NPM and has rapidly risen to the role of absolute protagonist in the field of study. Gradually, it has shown promise for actuation, catalysis, energy, and sensing-related applications, just to cite a few \cite{wittstock2009nanoporous,wittstock2011nanoporous,fujita2012atomic,Yan2012Nanoporous,detsi2013metallic,kramer2004surface,xue2014switchable,qian2007surface,biener2008nanoporous,grosu2020hierarchical}.

A key enabling technology issue stands, nonetheless, in the way to the practical exploitation of NP Au, and NPM in general, namely their mechanical properties \cite{volkert2006approaching,kelly1991brittle,corcoran1992porosity,friedersdorf1996film,senior2006synthesis,hodge2006characterization,jin2018mechanical,richert2020review,soyarslan20183d,zandersons2021factors}. Actuators, self-supported catalysts and scaffolds for surface-enhanced Raman spectroscopy, thermal insulators and structural materials all need long-term mechanical stability and, therefore, a satisfactory characterization of the NPM response to elastic and plastic deformation. However, the subject is also per se extremely interesting. Porosity degree and characteristic lengths severely affect the response to mechanical loads. Particularly intriguing is the evidence that the reduction of ligament and pore size to a few nanometers can induce a significant hardening and stiffening of NP architectures. Although literature data are still fragmentary and contradictory, there is no doubt, at least for NP Au, that size effects tune somehow the mechanical properties.

Concerning the elastic properties in particular \cite{pia2018nanoporous,soyarslan20183d,mathur2007size,roschning2016scaling,shi2021scaling,ngo2015anomalous}, experiments and computer simulations have been used to investigate Young's modulus and Poisson ratio. Constitutive models have been mostly based on the Gibson-Ashby scaling laws \cite{gibson1982mechanics,gibson2003cellular}, but their validity has been questioned. Typically, ligaments have aspect ratios far from those compatible with Gibson-Ashby’s models, and other structural and topological factors have been shown to play a role \cite{mathur2007size,liu2017scaling}. For instance, several experiments have provided robust evidence that the NP Au Young's modulus increases as the ligament thickness is reduced approximately from 200 nm to 5 nm \cite{pia2018nanoporous,mathur2007size,biener2006size}. Surface stress has been invoked to explain experimental observations as well as the enhanced moment of inertia of finer structures \cite{mathur2007size} and the content of lattice defects \cite{wang2017mechanical}. In this latter case, it is worth recalling that, while point defects can induce a significant reduction of the elastic moduli \cite{jhi2001vacancy,duncan2006role}, extended defects may affect local interatomic bonding \cite{wang2017mechanical}.

Motivated by this intriguing and open scenario, in this work we investigate the Young's modulus of NP Au combining experimental, simulative and theoretical methods. To this aim, we have generated in silico replicas of NP Au structures with atomistic resolution by mapping the initial volumetric data obtained by experimental tomographic evidence. Re-scaling the characteristic lengths of the atomistic sample, we have systematically explored the impact of size effects, being the porosity degree the same.

Our findings confirm that the Young's modulus significantly increases as the ligament size decreases. Further, we provide evidence that the observed stiffening cannot be ascribed to surface stress or grain-boundary effects. Rather, our results point out a quantitative correlation with the dislocation density. Specifically, we show that local stress and strain fields in the neighbourhood of dislocation cores allow dislocations to work as reinforcing solutes.

\section{Results and discussion}
\subsection*{3D reconstruction and rendering of NP Au samples.}
We used serial block-face (SBF) scanning electron microscopy (SEM) to collect two-dimensional images of NP Au samples through a SEM-dedicated VolumeScope cutting device (see Materials and Methods for details). A typical SEM image is shown in Fig \ref{fig:1new} (Panel a) for illustration purposes. The typical appearance of a NP architecture is immediately evident, with irregularly shaped ligaments and ligament junctions forming condensed open cells finally resulting in an inherently distorted three-dimensional lattice.
The two-dimensional images were assembled into a single three-dimensional structure via a suitable volume reconstruction method (see Materials and Methods for details). A representative tomographic reconstruction of the NP Au structure is shown in  in Fig. \ref{fig:1new} (Panel b). It allows clearly appreciating how solid and void give rise to a maze of rounded shapes, where ligaments interconnecting at massive junctions form an interpenetrating, bi-continuous network of pores and ligaments and nodes. Ligaments and pores have similar diameter of about 42 $\pm$ 3 nm. Porosity evaluation throughout the reconstructed sample volume indicates a solid volume fraction of 0.48 $\pm$ 0.03.
 
%Serial block-face (SBF) scanning electron microscopy (SEM) images have been collected by using 
 %a SEM-dedicated VolumeScope cutting device and then assembled into volume files (see Section Materials and Methods for a detailed description). By using the reconstructed volume, the porosity (i.e, the volume of the pores) was also determined slice by slice over the whole thickness of the 3D-imaged samples.  A representative image of a typical dealloyed nanoporous gold sample is shown in Fig.\ref{fig:1new} (Panel a). Solid and void regions give rise to an elegant maze of rounded shapes. Ligaments interconnect at irregular, massive junctions forming a disordered assembly of open cells with various, distorted geometries. Overall, this architecture consists of an interpenetrating, bi-continuous network of pores and ligaments and nodes. In average, ligaments have diameter equal to 42 $\pm$ 3 nm and pores have comparable size. The combination of geometrical dimensions and mass measurements results in a solid volume fraction of 0.48 $\pm$ 0.03. 
\subsection*{``In silico'' sample generation with atomistic resolution}
\begin{figure}[t]
\includegraphics[width=1\textwidth]{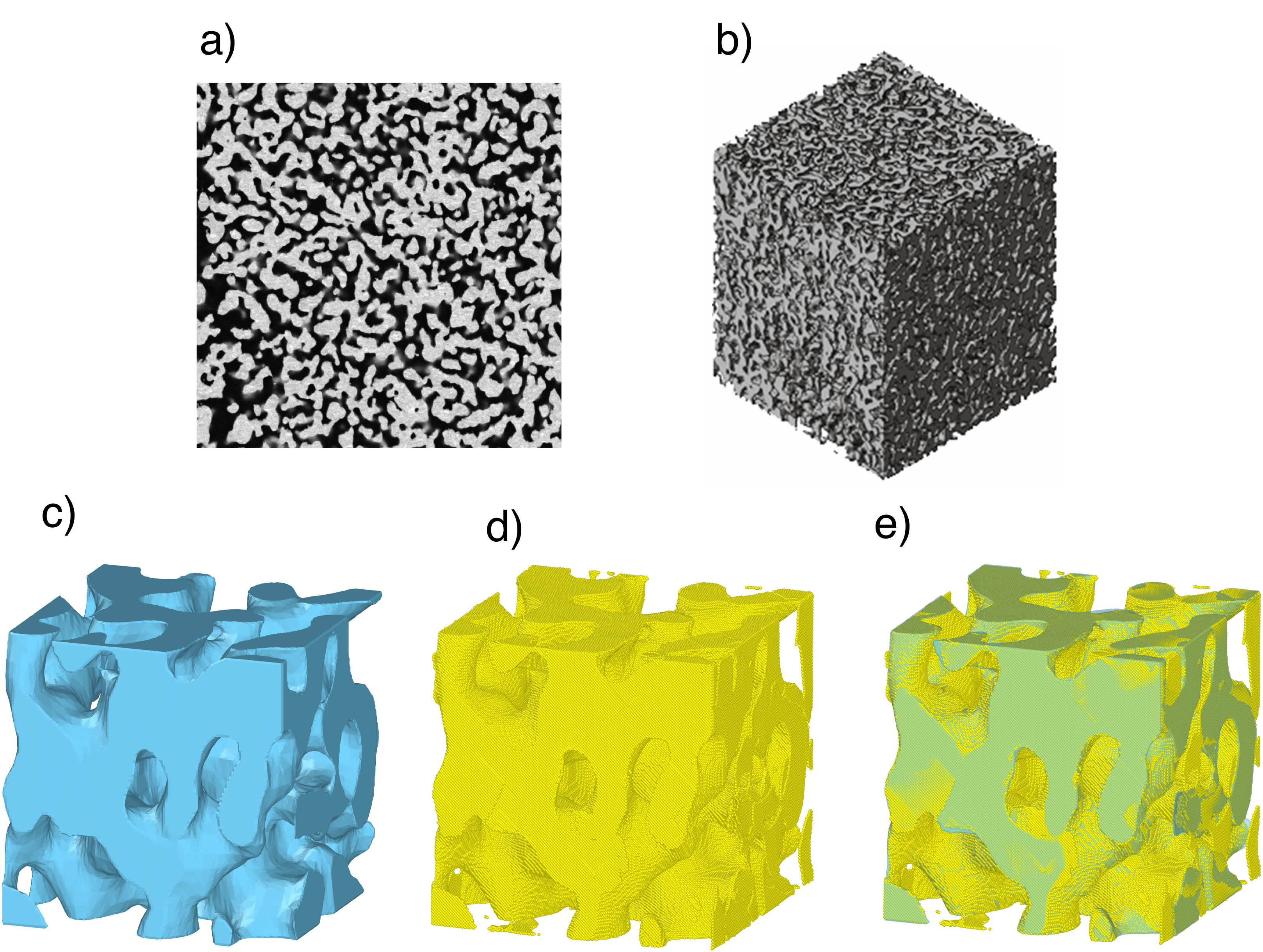}
\caption{Panel (a), a typical SEM two-dimensional image of an NP Au sample. Panel (b), representative tomographic reconstruction of an NP Au structure. Panel (c), cyan color: volumetric structure of an NP Au sample with $L_{cell}$=19.7 nm. Panel (d), yellow color: corresponding structure obtained by using the ``in silico'' procedure described in the main text. Panel (e): superposition of the volumetric and atomistic pictures, proving the almost perfect matching between the two structures.%Structure of nanoporous gold (sample dimension: 20.1$\times$19.6$\times$20.0 nm$^3$). 
}
\label{fig:1new}
\end{figure}

Starting from the volumetric data previously obtained for the NP Au samples, we proceeded to decorate the corresponding SBF-SEM images with an atomistic structure corresponding to face-centered-cubic (fcc) Au lattice with spacing $a$= 0.40782 nm. The procedure has been performed by following 3 consecutive steps:

\begin{enumerate}
\item Homotetic transformation - We used the Meshmixer code \cite{schmidt2010meshmixer} to apply a series of homothetic transformations to the selected experimental cubic NP Au sub-volumes. Preserving shape and inner geometries while changing the volume, we generated 9 NP Au samples with side length $L_{cell}$ ranging from 5.5 to 40.7 nm per each selected sub-volume. Fig. \ref{fig:1new} (Panel c) shows the NP Au sample with $L_{cell}$=19.7;
%starting from the experimental cubic structure with lateral dimension of 200 mm, we applied a series of homotetic transformations (which preserve the system shape by modifying the corresponding volume) by reducing the system lateral dimension down to the nm length-scale. In particular, we generated 9 samples 
 %lateral dimensions raging from 15 up to 40 nm. This procedure has been performed using the Meshmixer code \cite{schmidt2010meshmixer}.
\item Solid phase identification – We used the Nanosculpt software \cite{prakash2016nanosculpt} to identify the NP Au volume occupied by the solid phase, and build the surface mesh that encloses it;
\item Crystallographic information addition - We filled the selected volume with Au atoms arranged in a perfect fcc lattice. The NP Au samples we obtained had number of atoms ranging from 4.95$\times$10$^3$ up to 1.85$\times$10$^6$. 

One out of the 9 cubic NP Au samples with $L_{cell}$=19.7 nm we have generated is shown in Fig. \ref{fig:1new} (Panel d). The comparison with the corresponding experimental SBF-SEM tomographic reconstruction reported in Fig. \ref{fig:1new} (Panel e) clearly shows how accurate is the atomistic decoration in reproducing the characteristic shapes of NP Au samples.
%Identification of the material phase within the nanoporous structure: we used the program Nanosculpt \cite{prakash2016nanosculpt} to identify the volume of the systems to be decorated with Au atoms and to define a surface mesh to enclose this volume. 
%\item Crystallographic information addition – We filled the selected volume with Au atoms arranged in a perfect fcc lattice. The NP Au samples we obtained had number of atoms ranging from 4.95$\times$10$^3$  (for the system with lateral dimension of 15 nm) up to 1.85$\times$10$^6$ (for the system with lateral dimension of 40 nm)

% Adding crystallographic information: we filled the previously selected volume with Au atoms by using the information on the orientation, structure and lattice parameter of individual crystallites.
%%\item Filling the identified volume with atoms: once the crystallographic data have been obtained, we proceed to decorate the specific volume previously determined with the a fcc structure. In detail, the specific volume is completely filled with Au atoms in the defined crystallographic orientation by rotating the primitive lattice vectors of the desired lattice structure into the given orientation and then placing atoms at all integer multiples of these lattice vectors, starting from a chosen origin.
\end{enumerate}

%By using this protocol we generated 9 different atomistic NP Au structures as shown in  Fig.\ref{fig:1new} (Panel b), containing a total number of atoms ranging from 9.50$\times$10$^4$  (for the system with lateral dimension of 15 nm) up to 1.85$\times$10$^6$ (for the system with lateral dimension of 40 nm). In Fig.\ref{fig:1new} (Panel c) we compare the atomistic NPM structure obtained by means of the above "in silico" procedure with the corresponding experimental volumetric representation, obtained by assembling different SBF-SEM images. We observe a very accurate description of the overall morphological and topological features of the experimental samples, a result which stands for the robustness of the procedure and its effectiveness to generate trustworthy atomistic NP Au samples.

\subsection*{Young's modulus of atomistic NP Au samples}

\begin{figure}[t]
\includegraphics[width=1\textwidth]{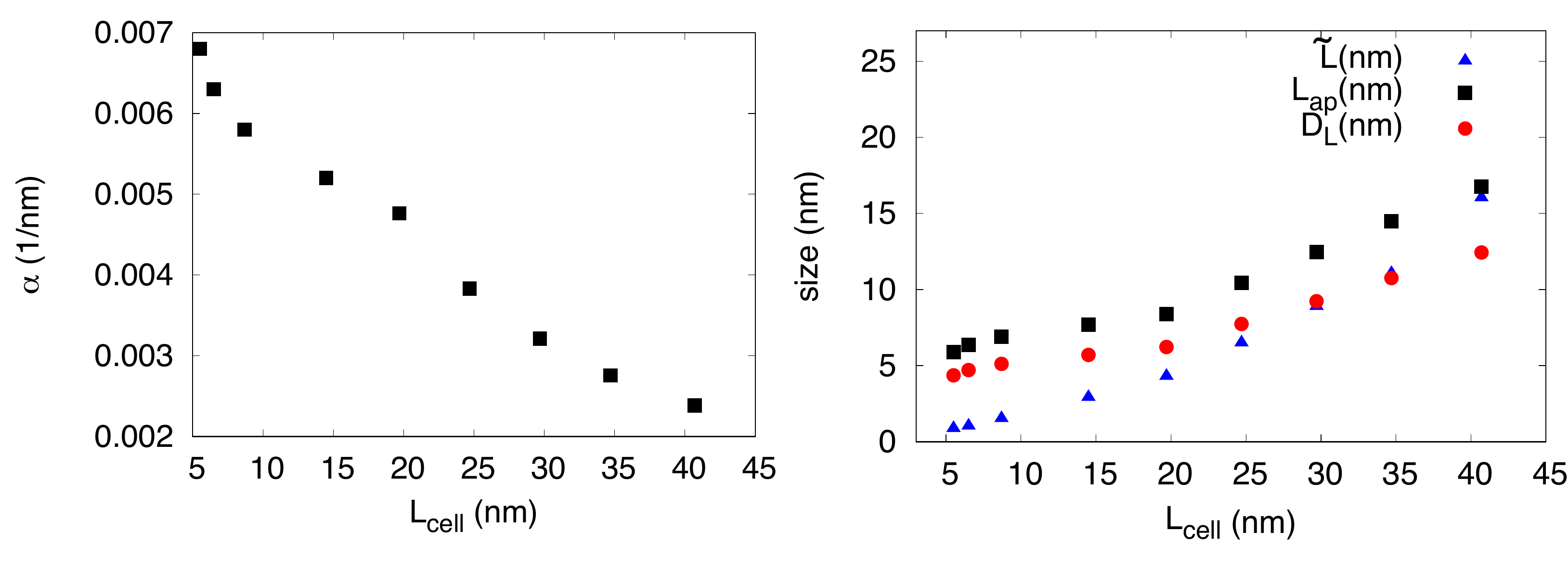}
\caption{Left: specific surface area $\alpha$ as a function of the side length $L_{cell}$ of the cubic NP Au sample. Right: apparent ligament diameter ($L_{ap}$), characteristic spacing between local centers of the solid or the pore space ($\tilde{L}$) and average ligament diameter ($D_{L}$) as a function of the cubic NP Au sample side length $L_{cell}$.}
\label{fig:2new}
\end{figure}

We used the Ovito software \cite{stukowski2009visualization} to evaluate the specific surface area $\alpha$ of the different NP Au samples. The corresponding values obtained from the 9 different NP Au sub-volumes selected per each NP Au sample size considered are plotted in Fig. \ref{fig:2new} (left) as a function of the side length $L_{cell}$ of the cubic NP Au sample. It can be seen, in the range of the explored $L_{cell}$ values, that $\alpha$ monotonically decreases as function of $L_{cell}$. 
%The specific surface area is routinely measured for NP Au structures. The experimental estimate for our NP Au sample is ***. Also shown in Fig. 2a, it agrees remarkably well with the value numerically obtained for the atomistic NP Au sample with side length of *** nm.

The specific surface area can be used to estimate the apparent ligament diameter $L_{ap}$. The calculation is immediate if the structure is assumed to consist of long circular rods. Accordingly,
 \begin{equation}
 \label{L-AP}
 L_{ap}=\frac{4}{\alpha}
 \end{equation}
 Other estimates  can be obtained assuming that the NP Au samples have a periodic diamond-like cubic structure interconnected by cylindrical ligaments of diameter $D_L$ \cite{huber2014scaling,ngo2015anomalous}:
 \begin{equation}
 \label{D-LIG}
 D_L=\frac{1.63(1.25-\phi)[1.89+\phi(0.505+\phi)]}{\alpha}
 \end{equation}
where $\phi$ is the solid volume  fraction $\sim$0.49 for all samples.
A more quantitative topological measure, which involves an exact relationship with the specific surface area, is the characteristic spacing $\tilde{L}$ between the centers of neighbouring ligaments. Since ligaments and pores most often have the same characteristic length, $\tilde{L}$ can be also regarded as a measure of the ligament size \cite{li2019topology}. In particular, $\tilde{L}$ can be expressed as \cite{li2019topology,soyarslan20183d}:
 \begin{equation}
 \label{L-TILDE}
 \tilde{L}=\left(\frac{g V_{total}}{G}\right)^{1/3}
 \end{equation}
 where $G$ is the topological genus measuring the number of connections in a specific network, $g$ is the scaled genus representing the number of connections in a representative volume element having size $\tilde{L}^3$ and $V_{total}$ is the total volume of the sample \cite{soyarslan20183d}. The calculations of $G$ and $g$ have been performed using the open-source code CHomP \cite{mischaikow2014chomp} and Eq. (16) of Ref. $33$, respectively (see Supporting Information for details). 
 The estimated values of $L_{ap}$, $D_L$ and $\tilde{L}$  are compared with each other in Fig. \ref{fig:2new} (right) as a function of $L_{cell}$. The different ligament size estimates undergo a consistent change with the total sample volume. In fact, they increase with $L_{cell}$ up to a maximum value of $\sim$16 nm. Eventually, we decided to use $\tilde{L}$  as the characteristic length of the selected NP Au structures and to refer to it the Young's modulus estimates obtained by subjecting the atomistic NP Au samples to uniaxial compression (see Supporting Information for details). Performed within the elastic limit, the simulated uniaxial compression determines a gradual deformation of the NP Au samples. A typical stress-strain curve is shown in Fig. \ref{fig:2} (right). Almost perfectly linear, its slope allows to univocally evaluate the Young's modulus of the corresponding NP Au structure. The Young's modulus $E$ is plotted in Fig. \ref{fig:2}  as a function of $\tilde{L}$. It exhibits a clean dependence on the length scale. In particular, $E$ increases approximately from $18.0$ GPa to $25.3$ GPa as   $\tilde{L}$ decreases.
 
An experimental estimate of the Young’s modulus, obtained by nanoindentation of the NP Au sample (see Materials and Methods for details), is also shown in Fig. \ref{fig:2} (left), shaded area. It is equal to 20.8 $\pm$ 2.3 GPa, i.e. it is almost coincident with the value of  18.3 GPa estimated by atomistic simulations for large $\tilde{L}$ values. The agreement between the two estimates goes beyond the initial expectations, based on the numerous sources of difference that can bring experimental and numerical values far apart for almost any physical quantity. In this case, the observed agreement can be tentatively ascribed to the high purity of the experimental NP Au sample and to the low level of local stresses probably achieved by the low-temperature thermal annealing to which the NP Au sample has been subjected.
%Figure \ref{fig:2} shows the calculated Young's modulus $E$ as a function of the ligament diameters $D_L$ for several computer-generated NP Au samples (see Section Materials and Methods). 

%Fig.\ref{fig:2} clearly shows that the Young's modulus strongly depends on $D_L$. In particular, we observe that $E$ increases as $D_L$ decreases ranging from a value of $18.0$ GPa for $D_L$=11.4 nm up to a maximum value of $25.3$ GPa for $D_L$=4.1 nm. Such interplay between $E$ and $D_L$ has been previously observed in different NPM structures but never univocally interpreted\cite{mathur2007size,biener2006size,pia2018nanoporous}. 

As outlined in the Introduction, several mechanisms can been taken into account in order to explain the $E=E(\tilde{L})$ dependence reported in Fig. \ref{fig:2}. In the next Section we analyze in detail all of them, by focusing on the primary NP Au elements, namely the single ligament, which we mimic as Au fcc-nanowire with a constant diameter $D_{nw}$.

%Finally we observe an underestimation of the estimated Young's modulus for relatively large  $D_L$ values ($\sim 18.3$ $\pm$ 0.1 GPa) with respect to the experimental value (24.6 $\pm$ 0.16 GPa).

\section*{Discussion}
A  variation of Young's modulus similar to the one reported in Fig. \ref{fig:2} has been already observed in several cases. However, it has never been interpreted univocally \cite{mathur2007size,biener2006size,pia2018nanoporous}. In contrast, various factors have been invoked to rationalize the experimental and numerical evidence. In the following, we examine the different possible mechanisms that can be expected to play a role in determining the observed mechanical response of NP Au structures.

\subsection*{Effect of surface-stress on fcc-nanowires}
Free surfaces are made of low-coordinated atoms. Reduced coordination numbers imply a redistribution of electronic charge and potential energy, possibly resulting in a change of local elastic moduli near the surface region compared with bulk ones. These effects have been extensively reported for clusters, wires and films \cite{miller2000size,dingreville2005surface,lachut2012effect,melis2017surface}. While the general formalism for the theoretical description of surface mechanics is fully established \cite{gurtin1975continuum,gurtin1978surface,sharma2003effect,chen2006derivation,steigmann1999elastic,javili2013thermomechanics,melis2017surface}, the surface constitutive equation and its parameters are still a matter of investigation \cite{dingreville2007semi}. In particular, surface effects, and the resulting size-dependent elastic behaviour, of Au nanowires remain unexplored.

%As a matter of fact, the atoms closed to a free surface exhibit a reduced coordination inducing a redistribution of the electronic charge and then of the energy.  As a result, the local elastic moduli near the surface region may differ from those of the bulk. These effects have been largely observed and studied for clusters, wires and films\cite{miller2000size,dingreville2005surface,lachut2012effect,melis2017surface}. 
%While the general formalism to theoretically describe surface mechanics is fully established \cite{gurtin1975continuum,gurtin1978surface,sharma2003effect,chen2006derivation,steigmann1999elastic,javili2013thermomechanics,melis2017surface}, the surface constitutive equation and its parameters are still a matter of investigation\cite{dingreville2007semi}. In particular, the surface effects and the resulting size dependent elastic behavior of gold nanowires remain unexplored. 

\begin{figure}
\includegraphics[width=1.0\textwidth]{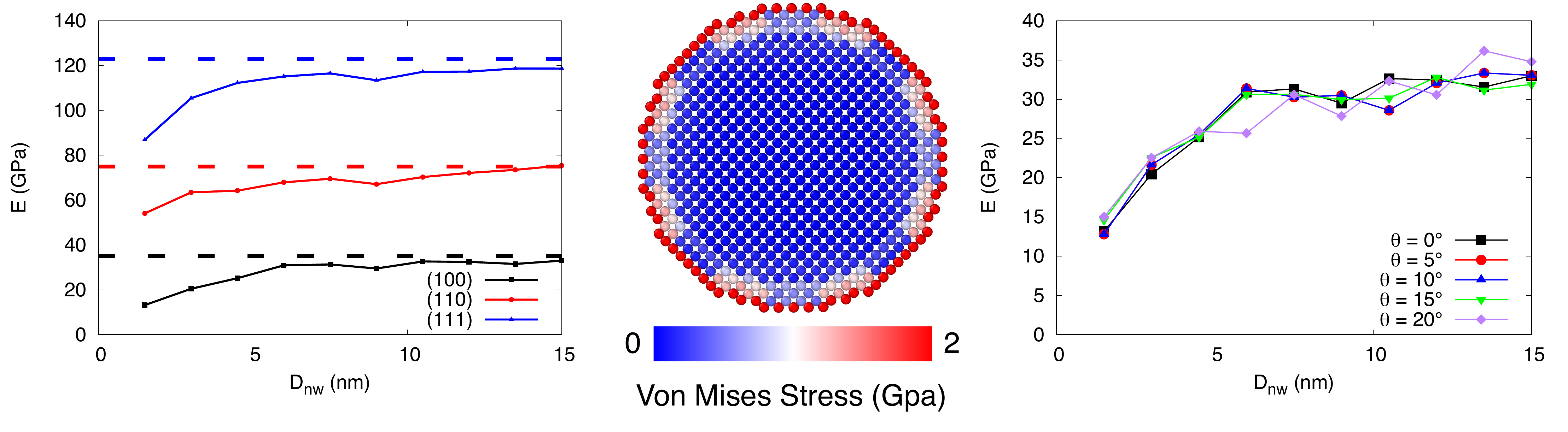}
\caption{Left: Young's modulus $E$ of gold-fcc nanowires with length $L_z$=20 nm oriented along the (100), (110) and (111) directions as a function of the diameter $D_{nw}$. Center: color map of the local von Mises stress calculated on a gold-fcc nanowires with $D_{nw}$=3 nm. Right: Young's modulus $E$ vs.  $D_{nw}$ of gold-fcc nanowires containing two grain-boundaries obtained by rotating by an angle $\theta$ the central part (half) of the NW with respect to the two top and bottom quarters.}
\label{fig:3}
\end{figure}

To elucidate this effect, we theoretically evaluated the Young's modulus of three macroscopic Au wires with their main axis aligned, respectively, to the three main crystallographic directions of the fcc lattice, namely (100), (110) and (111) (see Supporting Information for details). Being referred to macroscopic Au nanowires for which surface effects are negligible, the obtained values have to be compared with the Young's modulus of Au nanowires with diameter on the nanometer scale, which can be significantly affected surface-stress effects. To this aim, we performed molecular dynamics (MD) simulations on fcc Au nanowires with the main axis parallel to the (100), (110) and (111) directions.
We considered Au nanowires with total length $L_z$ of 20 nm and diameter varying in the range 1.5 nm $\le$ $D_{nw}$ $\le$ 15 nm. After a geometry optimization based on the conjugate-gradients algorithm, we obtained the Young's modulus values summarized in Fig. \ref{fig:3} (left), where $E$ is plotted as a function of the nanowire diameter $D_{nw}$ . 
The dashed lines account for the Young's moduli of macroscopic Au wires.

The Young's moduli of Au nanowires exhibits significant deviation from the macroscopic counterparts, decreasing down to 60\% as the nanowire size decreases below 10 nm. The observed reduction can be entirely ascribed to surface effects. This means that surface stresses induce a softening of the Au nanowire, not a stiffening. It follows that surface effects cannot explain the Young's modulus increase observed when the characteristic length of NP Au structure decreases. Fig. \ref{fig:3} (center) shows a color map of the local von Mises stress calculated on a Au-fcc nanowires with $D_{nw}$= 3 nm.

\subsection*{Effect of grain boundaries on fcc Au nanowires}
We applied the methodology described above to evaluate the Young's modulus of an Au nanowire containing two grain boundaries obtained by the rotation of the central region of the nanowire with respect to the end regions (see Supporting Information for details). In this way, grain boundaries are perpendicular to the nanowire main axis. The Young's modulus estimates are plotted in Fig.\ref{fig:3} (right) as a function of the nanowire diameter for the different rotation angles considered. It can be seen that the presence of grain boundaries determines a reduction of the Young's modulus as the nanowire diameter decreases. Therefore, similar to surface effects, grain-boundaries cannot justify the observed stiffening of NP Au structures with increasingly reduced ligament size.

In this regard, it is worth noting that the case study we examined is the most favourable to sizeably affect the elastic properties. For instance, let us consider the uniaxial compression of the fcc Au nanowire along its main axis. It is intuitively clear that grain boundaries perpendicular to the main axis have a minor impact compared with grain boundaries with a different orientation. In particular, oblique grain boundaries are likely to give rise to grain boundary sliding events upon compression that make the Au nanowire much softer and deformable. Similar behaviour has been reported for several metal nanowires \cite{schiotz2003maximum,zhu2019situ,saha2017investigation}. For this reason, we did not investigate further along this direction.

\subsection*{Effect of extended defects: dislocations}
Grain-boundaries and dislocations can be also expected to affect the elastic properties of NP metals \cite{patil2020hardening,dou2011deformation,ruestes2016hardening}. Many experimental works pointed out the effects of extended defects on the Young's modulus of nanowires \cite{dai2015elastic,chen2016effect,nam2006diameter,liu2006situ}. For instance, it has been shown that a high density of stacking faults in GaAs nanowires can determine a Young's modulus enhancement up to 13\% compared with the one of defect-free structures \cite{chen2016effect}.

\begin{figure}
\includegraphics[width=1\textwidth]{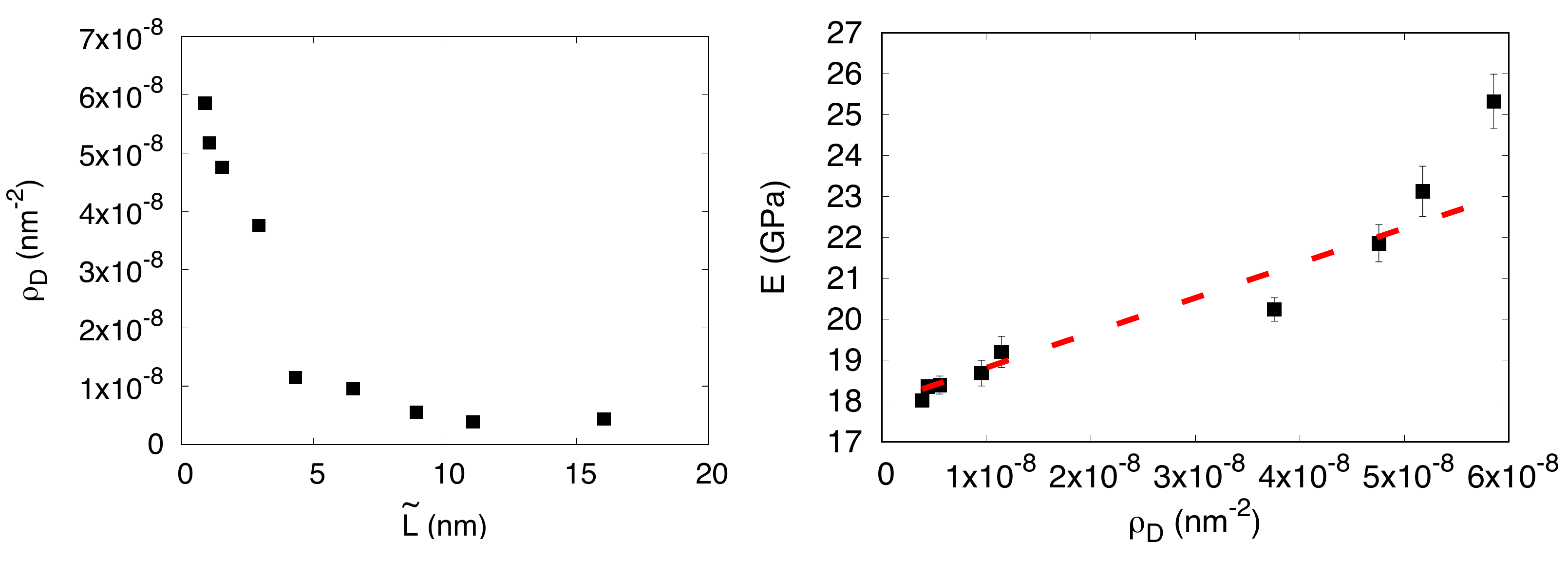}
\caption{Left: dislocation density $\rho_D$ as a function of the the characteristic spacing between local centers of the solid or the pore space $\tilde{L}$. Right: Young's modulus as a function of the dislocation density $\rho_D$ showing a reasonable linear dependence (red dashed line).}
\label{fig:4}
\end{figure}

Dislocations are also expected to affect the elastic properties of NP Au structures because of the associated stress and strain fields. We tried to identify the presence of dislocations in the 9 NP Au structures previously generated by means of the dislocation extraction algorithm \cite{stukowski2010extracting,stukowski2012automated}. Fig. \ref{fig:4} (left) shows the dislocation density $\rho_D$  as a function of the the characteristic spacing between local centers of the solid or the pore space $\tilde{L}$ calculated for the 9 NP Au samples under investigation. We clearly observe a sizeable $\rho_D$ reduction by increasing  $\tilde{L}$ suggesting that the occurrence of large $\rho_D$ values, observed for samples with relatively small  $\tilde{L}$, could in principle give rise to a stiffening effect, in turn resulting as an increase of the Young's modulus.  This is fully confirmed in Fig. \ref{fig:4} (right), where a reasonable linear correlation is observed between the Young's modulus and the corresponding $\rho_D$ value, indicating that high dislocation densities lead to a sample stiffening.

\begin{figure}
\includegraphics[width=1\textwidth]{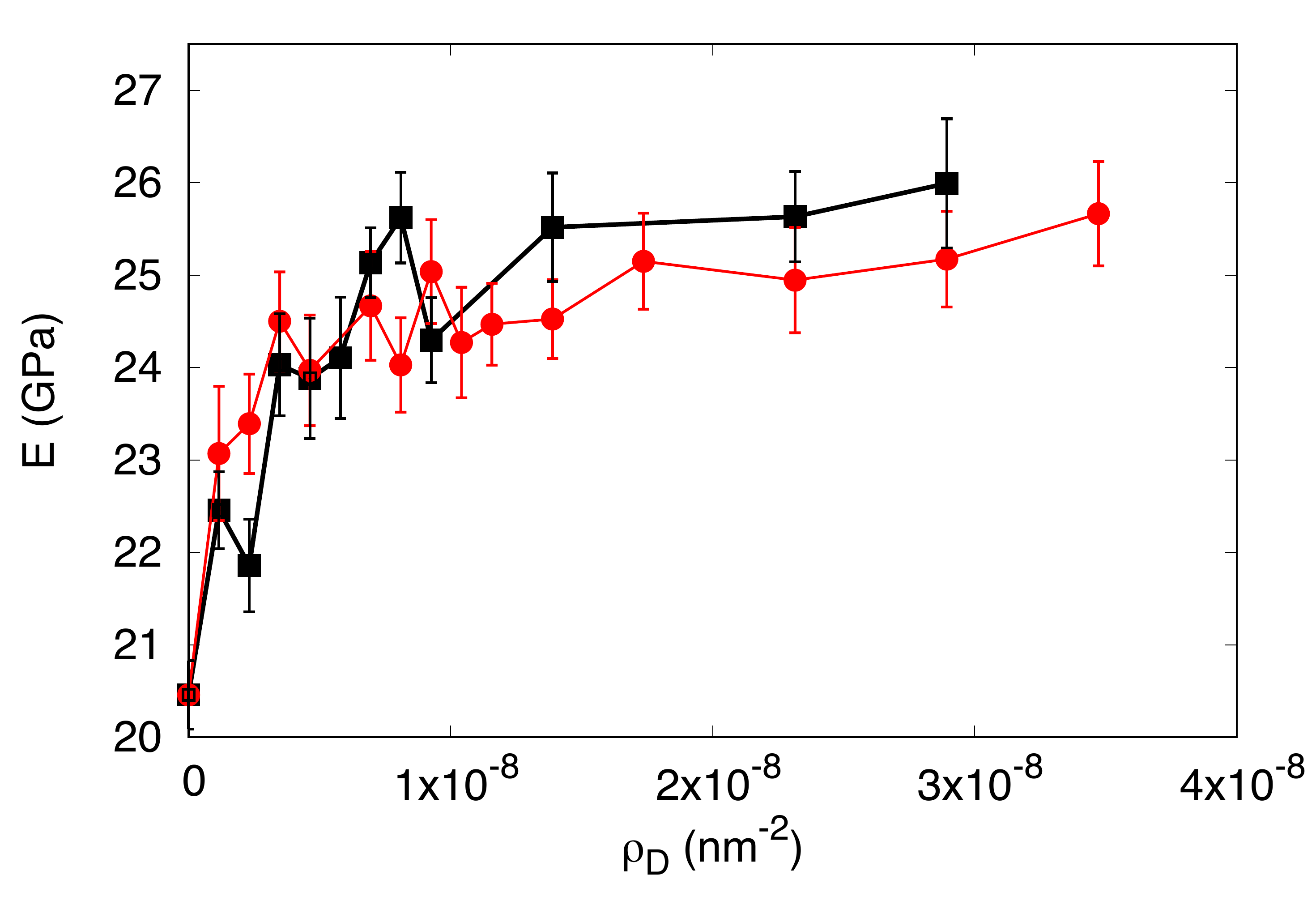}
\caption{Young's modulus $E$ of a gold fcc nanowire (oriented along the (100) direction with D$_{nw}$= 3 nm and L$_{z}$=20 nm) as a function of the edge-dislocation density $\rho_D$. Black dots and line: dislocations are positioned so as to maximize their distance. Red dots and lines: dislocations are placed in totally random positions.}
\label{fig:5}
\end{figure}

In order to make the above conclusion more robust, we  further investigated the effect of dislocations on the mechanical response of individual ligaments, by considering an Au nanowire with the main axis oriented along the (100) crystallographic direction, a diameter of 3 nm and a total length of 20 nm. Specifically, we evaluated the Young's modulus of the nanowire decorated with an increasing number of dislocations.
%Besides surface stress, extended defects such as grain-boundaries or dislocations could in principle affect the elastic properties of NPM. More specifically, many experimental works evidenced the role of extended defects on the the Young's  modulus of different   nanowires\cite{dai2015elastic,chen2016effect,nam2006diameter,liu2006situ}. In particular, in the case of GaAs nanowires, a  Young's modulus increase as large as 13\%  with respect to defect-free structures has been measured in the those containing a high density of stacking faults\cite{chen2016effect}.
%Dislocations are also expected to affect the elastic properties of NP Au structures because of the associated stress and strain fields. To estimate the effect of dislocations on the mechanical response of individual ligaments, we considered an Au nanowire with the main axis oriented along the (100) crystallographic direction, a diameter of 3 nm and a total length of 20 nm. Specifically, we evaluated the Young's modulus of the nanowire decorated with an increasing number of dislocations (see Supplementary Information for details).
Although different types of dislocations can be present in a typical NP Au sample, we restricted our investigation to edge dislocations, which can be easily generated in any specific position of the crystalline lattice \cite{groh2009dislocation}. We carried out calculations using two different dislocation arrangements. In one case, dislocations were suitably placed to maximize the distance between them. In the other, their positions were selected randomly. The obtained results are shown in Fig. \ref{fig:5}. In both cases, the Young's modulus $E$ undergoes a marked increase  up to $\sim$25\% for dislocation densities of $\sim$2$\times$10$^{-8}$ nm$^{-2}$, clearly showing that dislocations can induce a significant stiffening of NP Au ligaments.

\begin{figure}
\includegraphics[width=1.1\textwidth,angle=0]{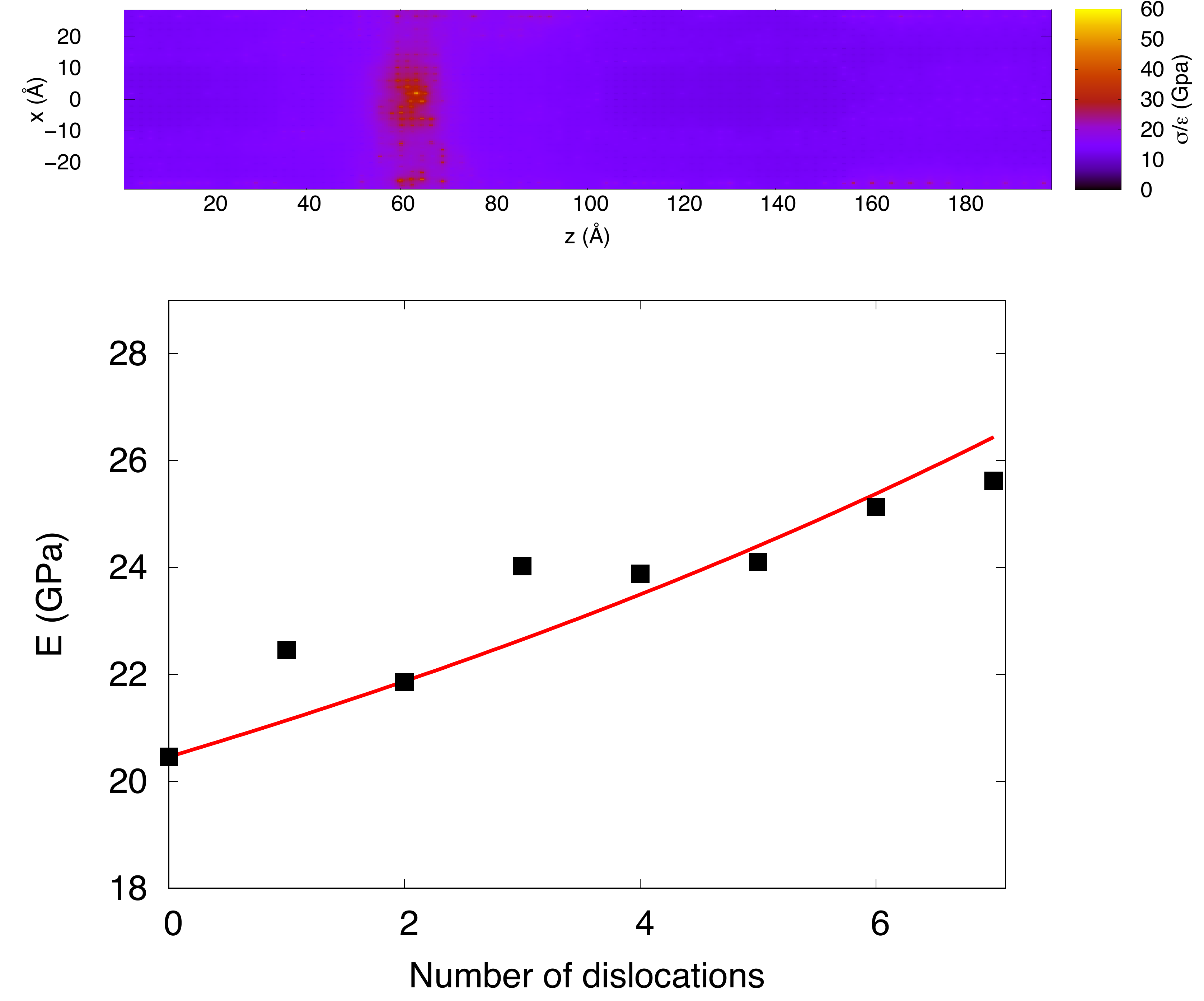}
\caption{Top: map of the Young's modulus, calculated as the ratio between the local stress and local strain values on a gold-fcc nanowire (oriented along the (100) direction with D$_{nw}$= 3 nm and L$_{z}$=20 nm) with a single edge-dislocation  positioned at $z$ = 6.3 nm. Bottom: black squares: Young's modulus $E$ of a gold fcc nanowire (oriented along the (100) direction with D$_{nw}$= 3 nm and L$_{z}$=20 nm) as a function of the number of edge-dislocations. Red line: effective Young's modulus obtained from Eq. \ref{eff} by considering $L=20$nm, $\ell_d/L\simeq 1/10$, $s_d/S\simeq 1/2$, $E=20.5$ GPa and $E_d=40$ GPa.}
\label{fig:6}
\end{figure}

We investigated the stiffening effect further estimating the local Young's modulus of a fcc Au nanowire containing a single edge dislocation (see Supporting Information for details). To this aim, we calculated the per-atom stress tensor and the corresponding per-atom elastic strain tensor. Their ratio provides a local measure of the Young's modulus. The map of local Young's modulus in a 1-nm-thick nanowire slab is shown in Fig. \ref{fig:6} (top). The Young's modulus is approximately equal to 20.5 GPa relatively far from the dislocation, but it increases up to 40 GPa nearby the dislocation. This is a clear evidence that dislocation effects stem from the local enhancement of elastic properties.
\subsection*{Theoretical considerations on dislocation effects}
Based on the numerical evidence concerning individual dislocations, we developed an elastic model to estimate the effects of multiple dislocations. We considered a nanowire with $n$ edge dislocations generating a surface $s_d$ over the total wire section $S$. We supposed that the effective longitudinal width of each dislocation is $\ell_d$ over the total wire length $L$. Additionally, we assumed that the effective volume of the dislocation $\ell_ds_d$ corresponds to an inhomogeneity with Young's modulus $E_d>E$. A procedure of elastic homogenization \cite{giordano2013analytical,colombo2011nonlinear} of the nanowire elastic response with the population of dislocations eventually leading to the effective Young's modulus as
\begin{eqnarray}
\frac{1}{E_{eff}}=\frac{1}{E}\left( 1-n\frac{\ell_d}{L}\right)+\frac{1}{E_d\frac{s_d}{S}+E\left( 1-\frac{s_d}{S}\right) } n\frac{\ell_d}{L},
\label{eff}
\end{eqnarray}
Eq. \ref{eff} is correct for $n\ll L/\ell$ as far as interactions between the elastic fields of nearby dislocations are negligible.

%Based on these results obtained for a single dislocation, we developed an analytical model able to to estimate the effect multiple dislocations. In detail, we consider a nanowire with $n$ edge dislocations generating a surface $s_d$ over the total wire section $S$. We suppose that the effective longitudinal width of the dislocation is $\ell_d$ over the total wire length $L$. Finally, we assume that the effective volume of the dislocation $\ell_ds_d$ corresponds to a reinforcing inhomogeneity with Youn g modulus $E_d>E$. A simple elastic homogenization of the elastic response of the nanowire with the population of dislocations eventually predicts the effective Young's modulus
%\begin{eqnarray}
%\frac{1}{E_{eff}}=\frac{1}{E}\left( 1-n\frac{\ell_d}{L}\right)+\frac{1}{E_d\frac{s_d}{S}+E\left( 1-\frac{s_d}{S}\right) } n\frac{\ell_d}{L},
%\label{eff}
%\end{eqnarray}
%which is correct for $n\ll L/\ell$ in order to avoid interactions between the elastic fields of the dislocations.

Theoretical predictions are compared with numerical findings in Fig. \ref{fig:6} (bottom). Data refer to an Au nanowire with total length $L$=20 nm and diameter $D_{nw}$=3 nm, dislocation width $\ell_d$=2 nm, surface associated with dislocations $s_d$=14.1 nm$^2$, bulk Au Young's modulus $E$=20.5 GPa and locally enhanced Young's modulus $E_d$=40 GPa. It can be seen that homogenized  elastic theory satisfactorily best fits the atomistic data, thus supporting the hypothesis that edge dislocations act as reinforcing dispersoids in the Au nanowire.
%5A direct comparison with MD simulations results can be drawn by considering $L=20$nm, %$\ell_d/L\simeq 1/10$, $s_d/S\simeq 1/2$, $E=20.5$GPa, $E_d=40$GPa. The good agreement can be observed in Fig.\ref{fig:6} (bottom) for $n=1,...,7$, confirming the reinforcing character of an edge dislocation in a gold nanowire. 
%For $n > 7$ we observe an overall saturation of the Young's modulus, which is not described by Eq.\ref{eff}, due to the interactions among dislocations (not included in the homogenized elastic model) which strongly reduce the reinforcing effect. 
We note that atomistic simulations involving more than 7 dislocations indicate that the Young's modulus reaches a plateau value of about 24 GPa. This can be ascribed to the fact that interactions among dislocations, not included in the homogenized elastic model, become increasingly more intense as the dislocation density increases, thus depressing the reinforcing effect.

The very good agreement between the atomistic data and the  analytical model further confirms the idea that dislocations act as reinforcing elements in single metal ligaments. This result ultimately explains the experimentally observed Young's modulus increases upon  reduction of the average metal ligament thickness.
\section{Conclusions}
In conclusion, we used experimental and theoretical methods to investigate the mechanical behaviour of nanoporous Au by addressing the long-standing debate of the observed Young's modulus increases upon characteristic length reduction.
To this aim we have used an experimental tomographic reconstruction to generate NP Au samples with atomistic resolution by mapping the initial volumetric data.
Atomistic simulations confirm that nanoporous Au stiffens as ligaments become finer, reproducing the experimental results obtained by nanoindentation.
A combination of  numerical and theoretical findings allow to rule out the effect of surface stress and grain boundaries.
In contrast, we have clear evidence that dislocations act as strengthening agents that enhance the nanoporous Au Young’s modulus.
The results obtained emphasize the importance of line defects and can address future research to investigate the mechanical properties of nanoporous metals. 
\section{Materials and methods}
\subsection*{Fabrication}
Nanoporous (NP) gold samples have been fabricated by electrochemical dealloying of a homogeneous chemically disordered polycrystalline Ag$_{70}$Au$_{30}$ solid solution. The parent alloy was prepared by mechanical alloying of Ag$_{50}$Au$_{50}$ powder mixtures followed by melting in an induction furnace and magnetic stirring.
Elemental Ag and Au powders with particle size below 45 $\mu$m and 99.99\% purity were manually mixed to obtain powder mixtures with Ag$_{50}$Au$_{50}$ stoichiometry. Mechanical alloying was carried out on powder batches of 10 g. The powders were sealed in a hardened-steel cylindrical reactor together with two 8-g stainless-steel balls under inert Ar atmosphere with oxygen and humidity contents below 2 ppm. The reactor was clamped to the mechanical arm of a SPEX Mixer/Mill 8000 ball mill and swung back and forth along a three-dimensional trajectory at about 875 rotations per minute. The mechanical processing resulted in the mutual dissolution of Ag and Au and the formation of a homogeneous chemically disordered nanocrystalline solid solution with Ag$_{50}$Au$_{50}$ chemical composition.

The obtained powders were placed in a quartz crucible and molten inside a laboratory Nabertherm N60/ER furnace. The melt was kept at 850 $\degree$C for 5 days under magnetic stirring and Ar flux conditions. Then, the molten phase was cast into cylindrical pellets about 1 mm high and with a diameter of about 1 cm. The surfaces of cylindrical pellets were finely polished and annealed for 10 h at about 400 $\degree$C K to relieve residual stresses.

Structural and microstructural transformations of powders and bulk solid along the different steps of the preparation process of the Ag$_{50}$Au$_{50}$ solid solution were investigated by X-ray diffraction (XRD). To this aim, a Rigaku SmartLab diffractometer equipped with a PhotonMax high-flux 9 kW rotating anode X-ray source was used. Not shown for brevity, XRD patterns were analyzed according to the Rietveld method. Mechanical alloying confirms its ability to induce the formation of homogeneous nanocrystalline solid solutions. Although we do not have direct evidence in this regard, melting and magnetic stirring can be expected to enhance the homogeneity of the final polycrystalline Ag$_{50}$Au$_{50}$ solid solution on the atomic scale.

NP Au structures were obtained from the parent alloy by electrochemically-assisted dealloying. The cylindrical pellets were exposed to an aqueous solution 0.75 M in HNO$_{3}$ and 0.01 M in AgNO$_{3}$. Selective Ag dissolution was carried out at room temperature inside a three-electrode electrochemical cell controlled by a Metrohm Autolab 302N potentiostat-galvanostat. A potential of about 600 mV was applied, using a Pt wire and an Ag electrode as counter and reference electrodes respectively.
Dealloying was interrupted after 10 days. To this aim, the cylindrical pellets were repeatedly rinsed with high-purity water and, finally, dried using an Ar flux at room temperature for 4 days. X-ray fluorescence measured a residual Ag content lower than 1 at \%.To eliminate internal residual stresses due to the dealloying process, the NP Au structures were subjected to a 5-day long annealing at 400 K under Ar flux conditions.
%A FEI Quanta 600 environmental scanning electron microscope (ESEM), working at an accelerating voltage of 10 kV and with a beam current of 1.7 nA mode, was used upon an in-chamber pressure of 300 Pa of water vapor for the study of NP gold isothermal coarsening. The microscope was equipped with a water-cooled heating stage, a high-temperature controller, and a K-type thermocouple for the thermal monitoring. A high temperature gaseous secondary electron detector (GSED) with a pressure limiting aperture (PLA), mounted on the bottom of the objective lens facing the specimen inserted in the heating stage, was used for the in situ ESEM imaging. The NP gold samples were fixed in a hollow graphite crucible, placed in the heating stage and heated from room temperature to 550 $\degree$C, 580 $\degree$C and 630 $\degree$C, respectively, at a rate of 20 $\degree$C min$^{-1}$ until 20 $\degree$C below the temperature set-point and then at a rate of 5$\degree$C min$^{-1}$ to reach the set-point minimizing the thermal drift. The temperature was then kept constant during the ESEM images acquisition.

\subsection*{Serial Block Face-SEM sectioning and imaging}
The NP gold sample was embedded in Epon epoxy resin using small silicon mold by polymerization at 65 $\degree$C for 72h. Resin blocks were mounted on aluminum specimen pins (10-006002-10, Micro to Nano) using cyan acrylic glue and trimmed with a glass knife to a rectangle 0.5 $\times$ 0.75 mm, with the gold sample exposed on all four sides. Silver paint (16 062-PELCO Conductive Silver Pain, TedPella) was used to favor the electrically grounding of the block edges to the aluminum pin. The entire, glued block was then sputter coated with a thin (25 nm) layer of gold (Cressington 208 HR sputter coater) to make its surface reflecting. After the block alignment into the SEM chamber, block cutting and serial block-face (SBF)-SEM images collection was performed using a dedicated VolumeScope (Thermo Scientific) cutting device mounted in a Teneo (Thermo Scientific) variable Pressure SEM, the latter operating at an acceleration voltage of 2 kV, and with a probe current of 50 pA and an in-chamber pressure of 20 Pa of water vapor. The sample was imaged at a magnification and with a pixel number that could allow obtaining an ultimate voxel size of 6 $\times$ 6 $\times$ 50 nm (x-y-z), where the z-one is the thickness chosen for the block cutting. The SEM images were collected using a dedicated backscattered electron (BSE) detector, provided for working with the VolumeScope cutting device.

\subsection*{3D reconstruction and rendering}
Serial SBF-SEM images were assembled into volume files using AVIZO software package (Thermo Scientific). NP gold ligaments and pores were segmented using Ilastik software30. The models extracted were assembled into volume and rendered using again AVIZO software package. By using the reconstructed volume the porosity (i.e, the volume of the pores) was also determined slice by slice over the whole thickness of the reconstructed sample.  

\subsection*{Nanoindentation}
The Young’s modulus of NP Au structures obtained by dealloying was measured using depth-sensing nanoindentation.  A calibrated three-sided pyramid diamond Berkovich tip with radius of about 200 nm was used. Indentation was carried out using a constant loading rate of 500 $\mu$N s$^{-1}$. The load was varied between 200 and 2000 $\mu$N. Up to 100 loading-unloading curves were measured per sample. The Young's modulus was estimated measuring the slopes of the linear part for any unloading curve and averaging over all unloading curves.
%The Young’s modulus of NP Au structures was measured using depth-sensing nanoindentation. A NanoTest Vantage Micro Materials indenter equipped with a calibrated three-sided pyramid diamond Berkovich tip with radius of about 200 nm was utilized. Indentations were carried out on the surface of cylindrical pellets applying loads between 200 and 2000 $\mu$N at the constant loading rate of 500 $\mu$N s$^{-1}$. Loading-unloading curves were measured on 40 indents per NP Au sample. Young’s modulus estimates were obtained from the slope of the unloading curve linear portion. Data were averaged over three NP Au samples.

\subsection*{Molecular dynamics simulations}
Once the atomistic NP Au structures have been generated, we performed a series of molecular dynamics (MD) simulations aimed at calculating their elastic properties as a function of the actual atomic scale structure. MD runs have been performed using the LAMMPS \cite{plimpton1995fast} simulation code and modelling the interatomic interactions by means of an embedded-atom model (EAM potential) \cite{foiles1986embedded} which has been previously validated as a good quantitative predictor of the elastic properties of gold systems \cite{ngo2015anomalous}. Atomic trajectories have been aged by the velocity-Verlet algorithm with a discretization of time evolution based on a time-step as small as 1.0 fs; temperature and pressure control was operated by a Nos\'e-Hoover thermostat and barostat, respectively. 

\begin{figure}[t]
\includegraphics[width=1\textwidth]{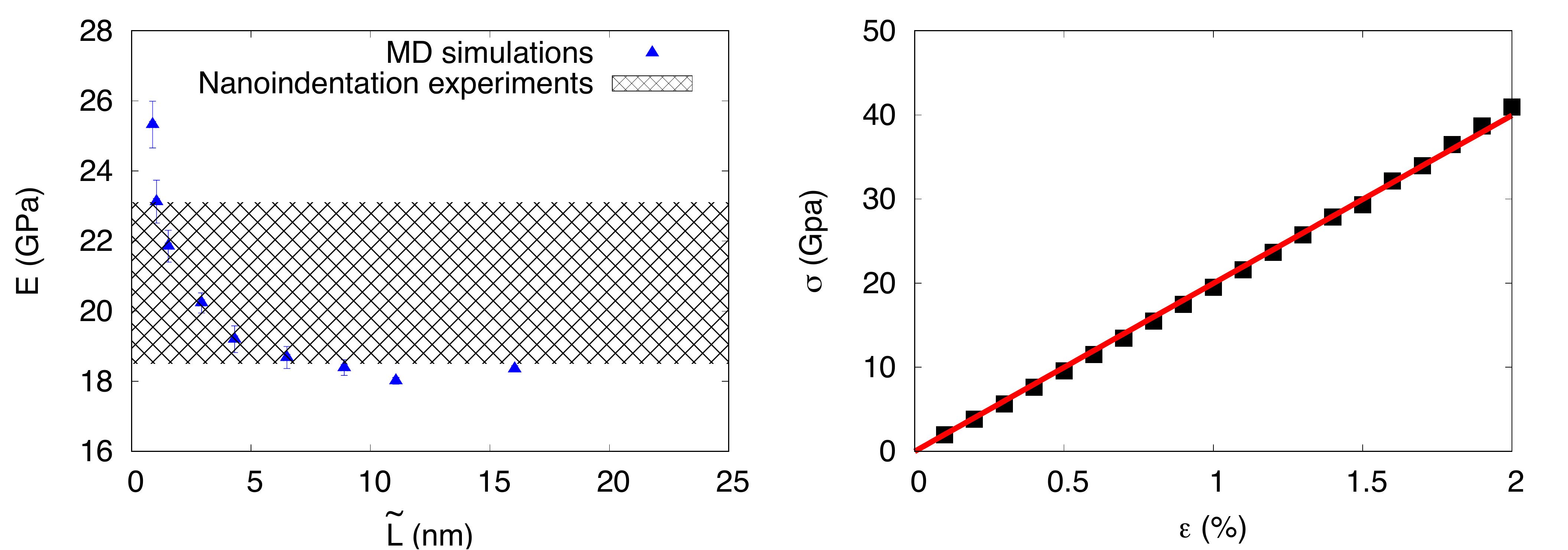}
\caption{Left: calculated Young's modulus $E$ for the computer-generated NP Au gold samples as a function of the characteristic spacing between local centers of the solid or the pore space $\tilde{L}$; the shaded black area represents the experimental estimate of the Young’s modulus obtained by nanoindentation. Right: a typical calculated linear elastic stress-strain curve (back squares) for a sample with $L_{cell}$= 40.7 nm; the Young's modulus is obtained by linear interpolation (thin red line).}
\label{fig:2}
\end{figure}

Upon ``in silico'' generation, NPM samples have been at first relaxed by performing a conjugate-gradient minimization, until the modulus of the maximum force on each atom was less than 10$^{-6}$ eV/\AA. The samples were then equilibrated at temperature T=300 K and pressure P=1 Atm for 100 ps and eventually cooled down to zero temperature by a gentle simulated annealing. After such a careful equilibration procedure, we statically applied a uniaxial tensile strain $\epsilon_{ii}$ (where $i$=$x$, $y$ or $z$)  on each NPM sample in the range $0\%\le\epsilon_{ii}\le2$\% at incremental steps as small as $\Delta\epsilon_{ii}= 0.1\%$. After another conjugate-gradient minimization of each deformed configuration, we calculated the resulting stress value $\sigma_{ii}$ by averaging the per-atom stress tensor \cite{thompson2009general} throughout the whole simulated structure.
Due to the relatively small maximum strain applied, we assume a regime of linear elastic response and we accordingly estimate the NPM Young's modulus by fitting the stress-strain curve. The inset of Fig.\ref{fig:2} shows a typical stress-strain curve with the corresponding fitting line obtained for a cubic NPM Au sample with 40.7 nm lateral size.
\begin{acknowledgement}
We acknowledge financial support by ``Fondazione di Sardegna'' under project ADVANCING (ADVAnced Nanoporous materials for Cutting edge engineerING
), call 2018 for basic research projects.
\end{acknowledgement}


\bibliography{acs-achemso}

\clearpage

\clearpage

\clearpage

\clearpage

\clearpage

\clearpage

\clearpage

\clearpage

%\begin{figure}
%\includegraphics[width=1\textwidth]{7.pdf}
%\caption{Black squares: Young's modulus $E$ of a gold fcc nanowire (oriented along the (100) direction with D$_{nw}$= 3 nm and L$_{z}$=20 nm) as a function of the number of edge-dislocations. Red line: effective Young's modulus obtained from Eq.\ref{eff} by considering $L=20$nm, $\ell_d/L\simeq 1/10$, $s_d/S\simeq 1/2$, $E=20.5$GPa and $E_d=40$GPa. %Eq.\ref{eff}  do not properly describe the saturation of the effective Young's modulus with increasing values on $n$ (blue circles) due to the interactions among dislocations. 
%}
%\label{fig:7}
%\end{figure}
%\clearpage
\end{document}

\section*{S.1: Calculation of the  characteristic spacing between local centers of the solid or the pore space $\tilde{L}$}
A characteristic topological quantity used to characterize the microstructure of NP materials is the spacing between local centers of the solid or the pore space $\tilde{L}$, which can be interpreted as an alternative measure of the mean diameter of the NP Au ligaments  \cite{li2019topology}.
In detail, $\tilde{L}$ can be estimated \cite{li2019topology,soyarslan20183d} from the topological genus $G$ (i.e., the number
of connections in the NP Au microstructure), the scaled genus $g$ (i.e., the number of connections in a representative volume element having size $\tilde{L}^3$) and the total volume of the sample $V_{total}$ \cite{soyarslan20183d}  :
\begin{equation}
 \label{L-TILDE}
 \tilde{L}=\left(\frac{g V_{total}}{G}\right)^{1/3}
 \end{equation}
The calculations of $G$ have been performed by estimating, using the CHomP \cite{mischaikow2014chomp} open-source code, the corresponding Betti numbers $B_0$ and $B_1$, representing a characteristic topological invariant of the NP Au samples. In particular, $B_0$  is an estimate of the number of connected components while $B_1$ represents the number of  handles a the specific structure. The genus is obtained as $G = B_1$.
The calculation of the scaled genus $g$ have been performed as \cite{soyarslan20183d}:
\begin{equation}
 \label{g}
g=\frac{2\pi \alpha^3}{3\sqrt{3}}(1-\xi^2)e^{-\xi^2/2}
 \end{equation}
where $\alpha$ is the specific surface area and  $\xi$ a characteristic function of the volume fraction  $\phi$:
\begin{equation}
 \label{xi}
\xi(\phi)=\sqrt{2}erf^{-1}(2\phi-1)
 \end{equation}
 where $erf$$^{-1}$ is the inverse error function. 
\section*{S.2: Effect of extended defects: grain boundaries}
We investigate the effect of grain-boundaries on the Young modulus of single ligaments by positioning in  a single gold fcc nanowire (oriented along the (100) direction with  length of 20 nm)  
two grain-boundaries obtained by rotating by an angle $\theta$ the central part (half) of the nanowire with respect to the two top and bottom quarters. Fig. 4 (right) of the main text shows the estimated Young modulus as a function of $D_{nw}$ in the case of different $\theta$ values in the range $0^{\circ}-20^{\circ}$. As opposed to the case of dislocations, the presence of grain-boundaries does not affect the Young modulus for all the $\theta$ values in the range $0^{\circ}-20^{\circ}$.

\section*{S.3: Young modulus of a macroscopic gold wire}
We theoretically determine the Young modulus of a macroscopic gold wire aligned to the three principal directions (100), (110) and (111) of the cubic crystal. To do this, we assume to apply a stress of the form $\hat{T}=T_0\vec{n}\otimes\vec{n}$, where $\vec{n}$ can assume the values (100), (110) or (111) directions. It means that $\hat{T}\vec{m}=0$ if $\vec{m}\perp\vec{n}$, and therefore the lateral surfaces of the wire are supposed to be free. For the cubic structure the stress-strain relation can be summarized as 
\begin{eqnarray}
T_{ii}&=&(C_{11}-C_{12})\varepsilon_{ii}+C_{12}(\varepsilon_{11}+\varepsilon_{22}+\varepsilon_{33}),
\label{ss1}\\
T_{ij}&=&2C_{44}\varepsilon_{ij}\,\,\,(\mbox{if }i\neq j),
\label{ss2}
\end{eqnarray}
where $C_{11}$, $C_{12}$ and $C_{44}$ are the bulk elastic constants. It should be noted that
Eqs.\ref{ss1} and \ref{ss2} reduce to the constitutive relation for an isotropic solid
when $2C_{44}=C_{11}-C_{12}$. The embedded-atom model (EAM potential) potential used for the present atomistic simulations predicts $C_{11}$= 78.1 GPa, $C_{12}$= 66.2 GPa and $C_{44}$= 21.2 GPa, in good agreement with  experimental data \cite{neighbours1958elastic}.
By using Eqs. \ref{ss1} and \ref{ss2}, we determine the strain tensor $\hat{\varepsilon}$, and we evaluate the directional strain $\varepsilon_0=\vec{n}\cdot\hat{\varepsilon}\vec{n}$. This allows to calculate  the directional Young modulus as $E_{\vec{n}}=T_0/\varepsilon_0$.  The Young modulus for three differently-oriented wires has been eventually obtained in the following form
\begin{eqnarray}
E_{(100)}=\frac{(C_{11}-C_{12})(C_{11}+2C_{12})}{C_{11}+C_{12}},
\end{eqnarray}
  \begin{eqnarray}
E_{(110)}=\frac{4(C_{11}-C_{12})(C_{11}+2C_{12})C_{44}}{2C_{11}C_{44}+(C_{11}+2C_{12})(C_{11}-C_{12})},
\end{eqnarray}
 \begin{eqnarray}
E_{(111)}=\frac{3(C_{11}+2C_{12})C_{44}}{2C_{12}+C_{11}+C_{44}}.
\end{eqnarray}
The results (dashed lines of Fig. 4 of the main text) represent the asymptotic Young modulus values obtained for macroscopic $D_{nw}$ values  for which surface effects becomes less critical. 

\section*{S.4: Calculation of the local Young modulus}
\begin{figure}
\includegraphics[width=0.8\textwidth,angle =-90]{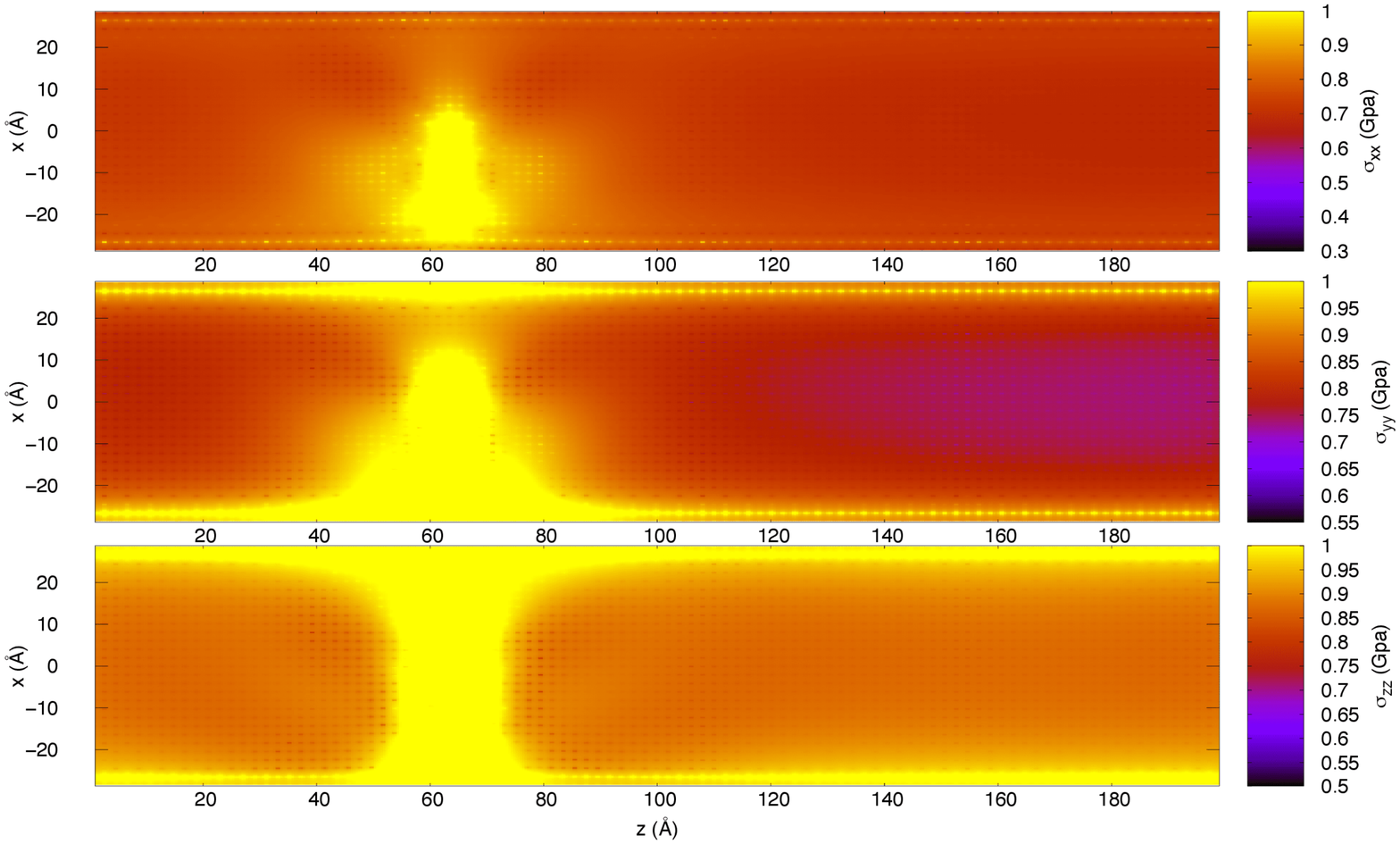}
\caption{map of the local stress tensor $\sigma^i_{jj}$ (where $jj$= $xx$, $yy$, $zz$), calculated on a gold-fcc nanowire (oriented along the (100) direction with D$_{nw}$= 3 nm and L$_{z}$=20 nm) with a single edge-dislocation  positioned at $z$ = 6.3 nm.} 
\label{fig:10}
\end{figure}

\begin{figure}
\includegraphics[width=0.8\textwidth,angle =-90]{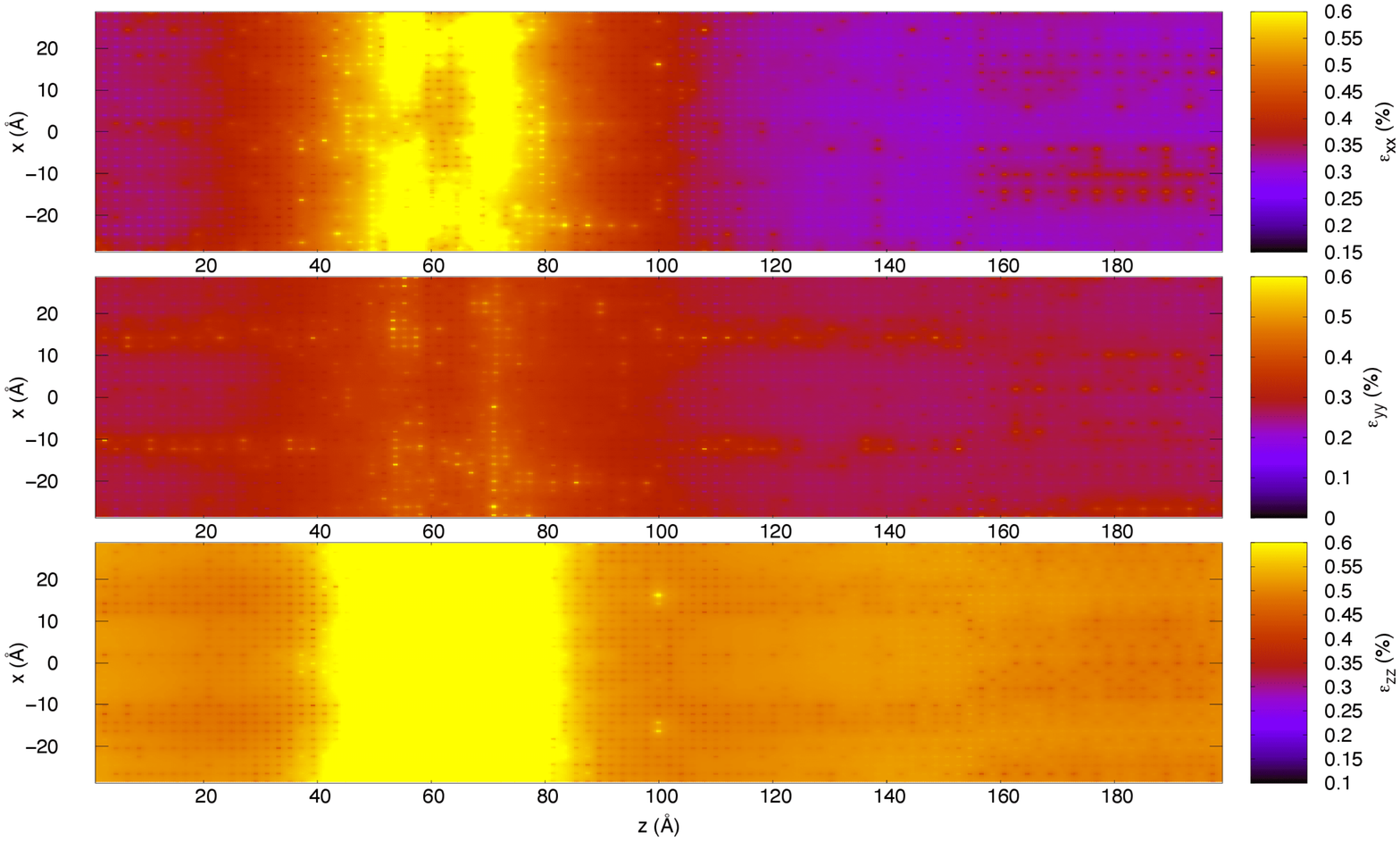}
\caption{map of the local strain tensor $\epsilon^i_{jj}$ (where $jj$= $xx$, $yy$, $zz$), calculated on a gold-fcc nanowire (oriented along the (100) direction with D$_{nw}$= 3 nm and L$_{z}$=20 nm) with a single edge-dislocation  positioned at $z$ = 6.3 nm.} 
\label{fig:11}
\end{figure}

\begin{figure}
\includegraphics[width=0.8\textwidth,angle =-90]{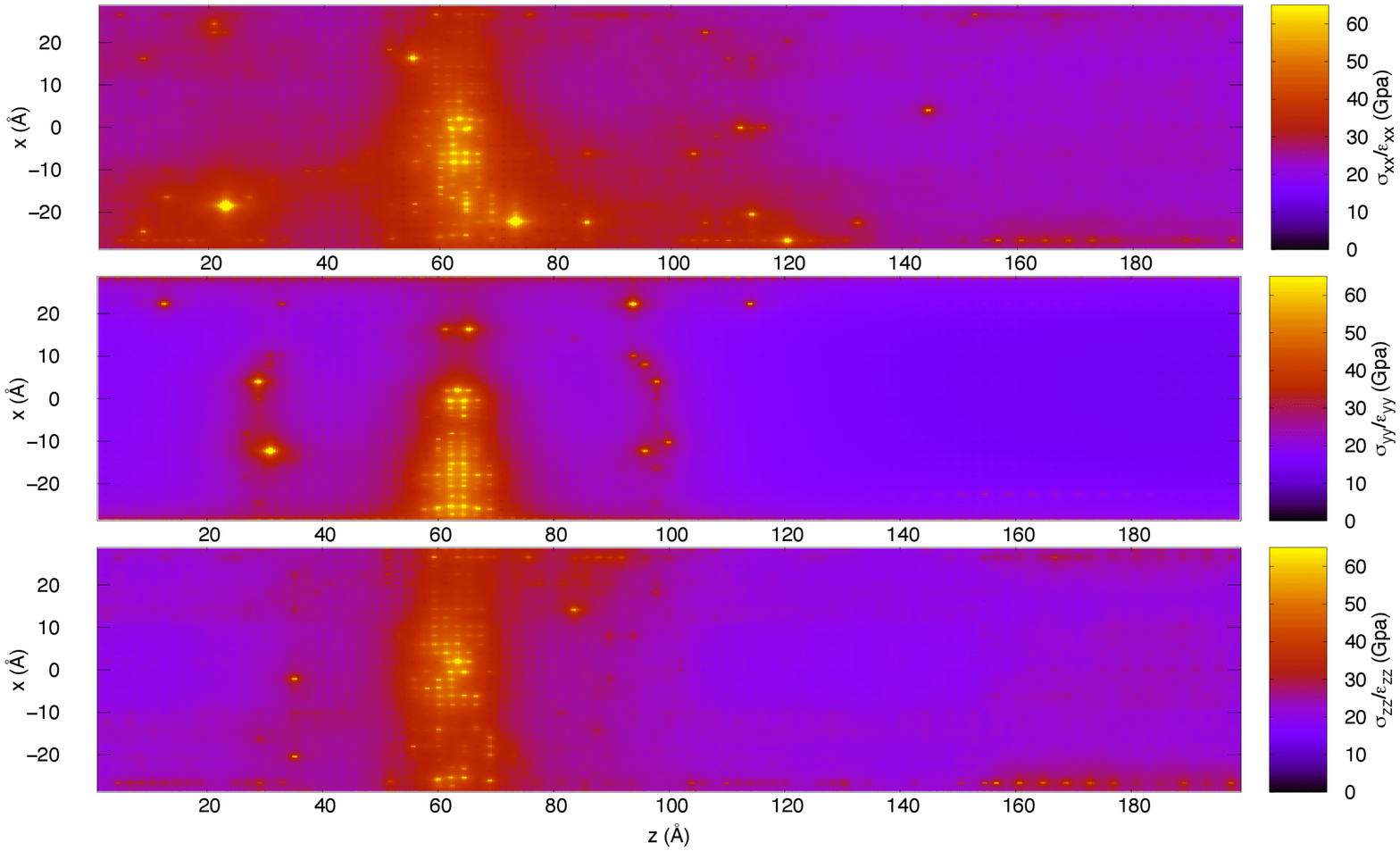}
\caption{map of the local Young modulus, calculated as the ratio between the local stress and local strain values (i.e., $\sigma^i_{jj}$/$\epsilon^i_{jj}$ where $jj$ = $xx$, $yy$, $zz$) on a gold-fcc nanowire (oriented along the (100) direction with D$_{nw}$= 3 nm and L$_{z}$=20 nm) with a single edge-dislocation  positioned at $z$ = 6.3 nm.} 
\label{fig:12}
\end{figure}

For each atom $i$ of a gold-fcc nanowire containing a single edge-dislocation positioned at z = 6.3 nm,  we estimated the the local Young modulus $E^{i}$ by calculating the ratio between the modulus of the  per-atom stress tensor $\sigma^i$ as implemented in the  LAMMPS package \cite{plimpton1995fast} and the corresponding per-atom elastic strain tensor $\epsilon^i$ as implemented in the Ovito package \cite{stukowski2009visualization}.
Fig. \ref{fig:10} shows a map of the local stress tensor components $\sigma^i_{jj}$ (where $jj$= $xx$, $yy$, $zz$) estimated on a nanowire slab having a thickness of 1 nm, while Figure \ref{fig:11} shows a map of the corresponding strain tensor components $\epsilon^i_{xx}$, $\epsilon^i_{yy}$, $\epsilon^i_{zz}$. In both cases we observe a significant  $\sigma$ and $\epsilon$ increase in the region (z $<$ 5.5  nm and z $>$ 7.5  nm) where the edge-dislocation was positioned. Figure \ref{fig:12} shows a map of the corresponding local Young modulus (i.e., $\sigma^i_{jj}$/$\epsilon^i_{jj}$ where $jj$ = $xx$, $yy$ and $zz$) showing a  sudden increase  up to a value $\sim$ 40 GPa in the region surrounding the dislocation confirming that the presence of dislocations locally increases the elastic properties of the ligaments. 
Fig. 6 in the main text shows the average value of the local Young modulus over the three $xx$, $yy$ and $zz$ directions.

%\clearpage
%\begin{figure}
%\includegraphics[width=1\textwidth]{8.pdf}
%\caption{Young modulus $E$ vs.  $D_{nw}$ of gold-fcc nanowires containing two grain-boundaries obtained by rotating by an angle $\theta$ the central part (half) of the NW with respect to the two top and bottom quarters.} 
%\label{fig:9}
%\end{figure}
\bibliography{acs-achemso}
\clearpage
\clearpage
\clearpage
